\documentclass[epsfig,graphics,twocolumn,floatfix,mathbbm,prl,nobibnotes]{revtex4}
\usepackage{graphicx}
\usepackage{amsmath}
\usepackage{amssymb}
\usepackage{subfigure}
\usepackage{gensymb}
\usepackage{braket}
\usepackage{bm} 
\usepackage{hyperref}

\begin{document}

\title{Multi-mode and long-lived quantum correlations between photons and spins in a crystal}

\author{Cyril Laplane}
\author{Pierre Jobez}
\altaffiliation{Present address: Institut f\"{u}r Experimentalphysik, Universit\"{a}t Innsbruck, Technikerstra{\ss}e 25, 6020 Innsbruck, Austria}
\author{Jean Etesse}
\author{Nicolas Gisin}
\author{Mikael Afzelius}
\email{mikael.afzelius@unige.ch}

\address{Groupe de Physique Appliqu\'ee, Universit\'e de Gen\`eve, CH-1211 Gen\`eve 4, Switzerland}

\begin{abstract}
The realization of quantum networks and quantum repeaters remains an outstanding challenge in quantum communication. These rely on entanglement of remote matter systems, which in turn requires creation of quantum correlations between a single photon and a matter system. A practical way to establish such correlations is via spontaneous Raman scattering in atomic ensembles, known as the DLCZ scheme. However, time multiplexing is inherently difficult using this method, which leads to low communication rates even in theory. Moreover, it is desirable to find solid-state ensembles where such matter-photon correlations could be generated. Here we demonstrate quantum correlations between a single photon and a spin excitation in up to 12 temporal modes, in a $^{151}$Eu$^{3+}$ doped Y$_2$SiO$_5$ crystal, using a novel DLCZ approach that is inherently multimode. After a storage time of 1 ms, the spin excitation is converted into a second photon. The quantum correlation of the generated photon pair is verified by violating a Cauchy - Schwarz inequality. Our results show that solid-state rare-earth crystals could be used to generate remote multi-mode entanglement, an important resource for future quantum networks.
\end{abstract}

\maketitle

\begin{figure*}[t]
\begin{center}
\includegraphics[width=0.95\textwidth]{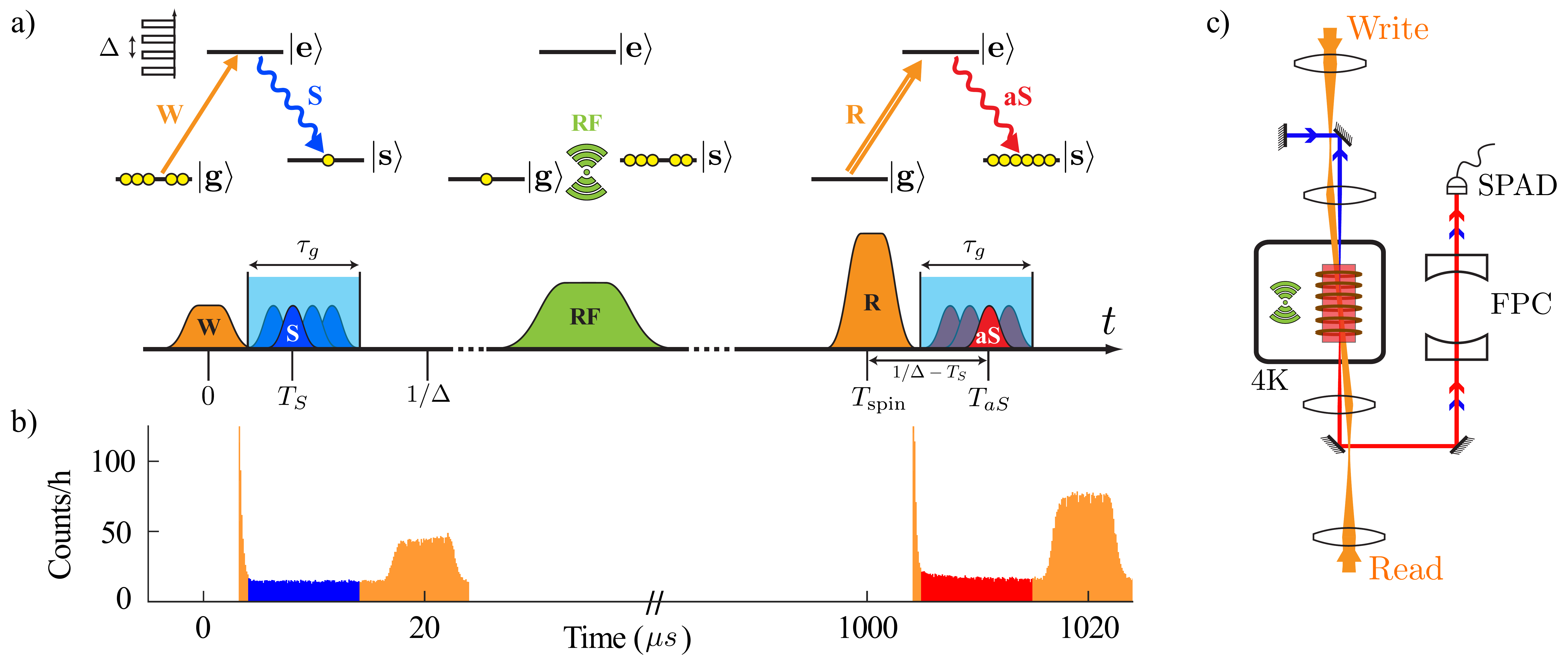}
\caption{The AFC-based DLCZ scheme and experimental set up. (a) We consider two spin ground states $\ket{g}$, $\ket{s}$ and an excited state $\ket{e}$ of the $^{151}$Eu$^{3+}$ ion in Y$_2$SiO$_5$. Initially $\ket{s}$ is emptied and a comb structure with periodicity $\Delta$ is prepared on the $\ket{g}\leftrightarrow\ket{e}$ transition using well-established optical pumping techniques \cite{Jobez2016}. A write pulse (W) resonantly excites a small fraction ($<1\%$) of the ions to $\ket{e}$ and the detection of a single spontaneous Stokes emission (S) heralds a single excitation in $\ket{s}$. A sequence of radio-frequency (RF) pulses resonant with the 34.5 MHz $\ket{g}\leftrightarrow\ket{s}$ spin transition extends the storage time to $T_{\mathrm{spin}}$ = 1000 $\mu$s \cite{Jobez2015}. The sequence contains an uneven number of pulses, which effectively inverts the population. A read pulse (R) excites the stored spin excitation to $\ket{e}$, which leads to a strong collective emission of an anti-Stokes photon (aS) at a time $T_{aS}$ such that $T_{S}+T_{aS} = T_{\mathrm{spin}} + 1/\Delta$. Ideally the read pulse should be a population inversion pulse covering the entire spectral bandwidth induced by the weaker write pulse. b) An example of a photon counting histogram showing the Stokes (blue region) and anti-Stokes (red region) detection gates. The pulses at 20 $\mu$s and 1020 $\mu$s are AFC echoes produced by the write and read pulses. c) The crystal is put in a low vibration cryostat at 4K. A coil is used to apply the RF sequence (in green). The write and read beams are counter propagating ($\mathbf{k}_W=-\mathbf{k}_{R}$), and have a small angle (about 3 $^{\circ}$) with respect to the Stokes/anti-Stokes modes ($\mathbf{k}_S$ and $\mathbf{k}_{aS}$). The phase matching condition $\mathbf{k}_W+\mathbf{k}_R-\mathbf{k}_S-\mathbf{k}_{aS} = 0$ then implies that $\mathbf{k}_{S}=-\mathbf{k}_{aS}$. A mirror retro-reflects the backward propagating Stokes/anti-Stokes modes, hence doubling the effective interaction length and allowing a single detection path consisting of a frequency-stabilized Fabry-Perot cavity (FPC) and a single photon avalanche diode (SPAD).}
\label{scheme}
\end{center}
\end{figure*}

Remote entanglement between light and matter is commonly based on the creation of quantum correlations between the internal state of a spin system and an optical mode \cite{Duan2001,Kimble2008,Sangouard2011}. This is due to the fact that spins can function as a long duration quantum memory for photons, which are ideal long-distance carriers of quantum information. Most schemes also rely on discrete quantum variables, i.e. a single spin excitation quantum correlated with a single photon propagating in the optical mode \cite{Sangouard2011}. The DLCZ approach is one of the most studied schemes for establishing such correlations, as it is based on relatively simple resources; an ensemble of atoms interacting with a coherent laser field \cite{Duan2001}. The ensemble of atoms is polarized into its lowest spin state $\ket{g}$, cf. Fig. \ref{scheme}a, then an off-resonant laser write pulse causes some spins to change their internal state to $\ket{s}$ through an optical Raman interaction. The creation of a spin excitation is strongly correlated to the emission of a corresponding Stokes photon in the optical mode.

The conventional DLCZ scheme has been widely implemented in atomic alkali gases at the single photon level \cite{Chaneliere2005,Eisaman2005,Yang2016}, but also for the entanglement of remote atomic ensembles \cite{Chou2005} as well as for the realization of entanglement swapping between two pairs of nodes \cite{Chou2007}. The scheme is, however, inherently single mode in time, hence only a single spin-photon correlation can be generated and stored. Spatial multimode generation can be achieved \cite{Lan2009}, but the time domain remains very attractive as it is compatible with propagation through low-loss single-mode fibers.

Solid-state devices are gaining interest as potential quantum nodes for quantum networks. A large variety of solid-state systems are currently being investigated, among those single NV centers \cite{Bernien2013}, quantum dots \cite{Delteil2016}, mechanical oscillators \cite{Riedinger2016} and ensembles of rare-earth ions \cite{Clausen2011,Rielander2014,Saglamyurek2015a,ZhouHuaLiuEtAl2015,Ferguson2016}. Rare-earth ion doped crystals are unique solid-state ensembles as they possess both long optical and spin coherence times, as well as large inhomogeneous linewidths in the optical domain \cite{Tittel2010b}. The combination of long optical coherence times (i.e. narrow homogeneous linewidths) and large inhomogeneous broadening implies that there is a large temporal multimode capacity \cite{Sangouard2011}. 

Quantum memory schemes based on inhomogeneous transitions, such as the gradient echo memory (GEM) \cite{Cho2016} and atomic frequency comb (AFC) memory \cite{Afzelius2009a} schemes, have been successfully implemented in order to store multiple temporal modes in a single spatial mode, in alkali gases \cite{Cho2016} and in rare-earth-ions doped crystals \cite{Jobez2016}, respectively. Yet, for DLCZ-type schemes, the use of inhomogeneous transitions for the generation of temporally multi-mode spin-photon correlations remains largely unexplored. Two recent experiments have demonstrated quantum correlations between light and spins in up to two temporal modes, in a laser-cooled $^{87}$Rb gas \cite{Albrecht2015} and a Pr$^{3+}$ doped crystal \cite{Ferguson2016}, respectively, with storage times limited to the range 10-20 $\mu$s. We also emphasize that in the latter experiment continuous variables states of light were used, which are extremely sensitive to propagation losses and require additional distillation techniques \cite{Ulanov2015}. Conventional DLCZ schemes use single photon states \cite{Duan2001}, however, as the loss of a photon does not increase the noise. 

Here we present experimental results on the generation of non-classical correlations between a single photon at 580 nm and a collective spin excitation in a $^{151}$Eu$^{3+}$:Y$_2$SiO$_5$ doped crystal. This specific rare-earth crystal features an optical coherence time that can reach 2.6 ms \cite{Koenz2003}, and a spin coherence time that can be stretched up to 6 hours \cite{Zhong2015} under very specific experimental conditions. A drawback of RE ions in general, however, and even more so of Eu$^{3+}$ ions in particular, are their weak oscillator strengths, which makes the off-resonant Raman excitation used in the DLCZ scheme very challenging to implement \cite{Goldschmidt2013}. Resonant excitation, on the other hand, introduces fast inhomogeneous dephasing of the generated collective spin excitation \cite{Ottaviani2009}, which leads to low read out efficiency of the anti-Stokes photons. A proposed solution to this dilemma is to employ inhomogeneous dephasing control on the optical transition \cite{Sekatski2011}, which in principle would allow efficient resonant generation of Stokes photons without reducing the memory read out efficiency. This can be done by creating an atomic frequency comb (AFC) on the optical transition, which is a technique that has been successfully employed to store quantum states of light on optical transitions \cite{Clausen2011,Rielander2014,Saglamyurek2015a,ZhouHuaLiuEtAl2015}.

\begin{figure*}[t!]
\includegraphics[width=0.95\textwidth]{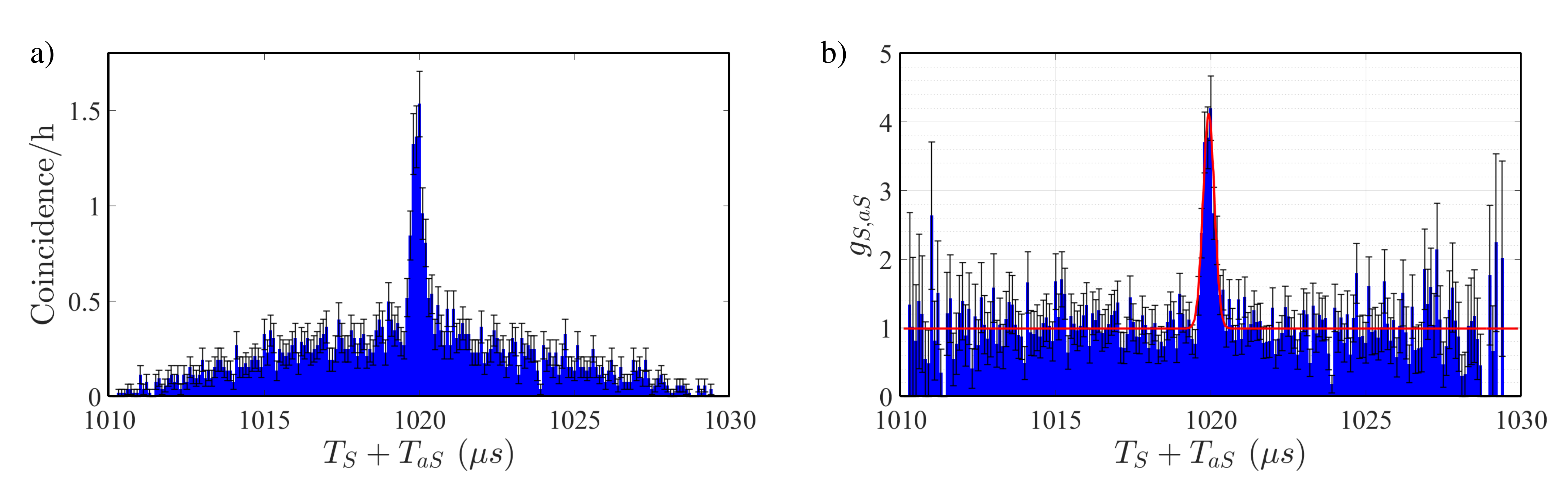}
\caption{(a) the Stokes/anti-Stokes coincidence detection histogram and (b) the normalized second-order cross-correlation function $g_{S,aS}$, as a function of $\tau=T_{S}+T_{aS}$. The Stokes and anti-Stokes detection gates were set to 10 $\mu$s and all detection events were binned into 100 ns. The coincidence histogram in (a) shows the expected peak at $\tau = T_{\rm spin}+1/\Delta$ = 1020 $\mu$s.  The cross-correlation peak in (b) is fitted using a Gaussian function (red curve), which results in full width at half maximum of 0.41 $\mu$s coherence time of the Stokes/anti-Stokes photon pair. This is in good agreement with the square excitation bandwidth of 2 MHz of the write pulse. If we define the duration of one mode as $2 \tau_c$, then the AFC-based DLCZ source generates correlated Stokes/anti-Stokes pairs in 10$\mu$s$/(2 \tau_c) = 12$ modes per experimental run.}
\label{gsi}
\end{figure*}

The main features of the AFC-based DLCZ scheme \cite{Sekatski2011} are shown in Fig. \ref{scheme}a. A resonant write pulse excites a part of the AFC comb on the $\ket{g}\leftrightarrow\ket{e}$ transition. The basic mechanism of the AFC comb is to periodically rephase any inhomogeneous dephasing induced by the absorption of an optical pulse, which leads to an echo emission after a time $1/\Delta$. In this scheme, however, the AFC echo of the write pulse is of no interest in itself. Instead, we consider the spontaneous emission of a Stokes photon on the $\ket{e}\leftrightarrow\ket{s}$ transition, at a time $T_S$ within a time window of duration $ \tau_g < 1/\Delta$. In Fig. \ref{scheme}b we present a typical photon counting histogram showing the Stokes detections following the write pulse. As the detection gate has a duration $\tau_g$ = 10 $\mu$s, which is much shorter than the radiative lifetime $T_1 = 1.97$ ms \cite{Koenz2003} of the excited state $\ket{e}$, the Stokes detection probability is constant over the entire gate.

By conditioning on the detection of a single Stokes photon, the ensemble of ions is projected into a single collective spin-wave excitation of the form

\begin{equation}
\ket{\psi_{SW}} \propto \sum_{j=1}^{N} e^{-i 2\pi \delta_j T_S} e^{i \mathrm{\Delta \textbf{k}} \cdot \mathrm{\textbf{r}}_j} c_j |g_1\cdot\cdot\cdot s_j \cdot\cdot\cdot g_N \rangle.
\end{equation}

\noindent Here $N$ is the total number of ions in the interaction volume ($N = \mathcal{O}(10^{12})$ in our case), $\delta_j$ is the optical detuning of ion $j$, $\mathrm{\Delta \textbf{k} = \textbf{k}_W - \textbf{k}_S} $ is the wave vector of the spin-wave, $\mathrm{\textbf{r}}_j$ is the position of ion $j$, and $c_j$ is the effective coupling constant of ion $j$. Importantly, the detection time $T_S$ is encoded in the amount of optical dephasing $\exp(-i 2\pi \delta_j T_S)$ that has occurred before the emission. This is the fundamental reason why the scheme is temporally multi-mode; the detection of time-separated Stokes photons leads to distinguishable spin waves.

In order to extend the storage time, we manipulate the spin waves using radiofrequency pulses (see Fig. \ref{scheme} and Supplemental Material for more details). Finally the stored excitation is mapped out by applying a read pulse on the $\ket{g}\leftrightarrow\ket{e}$ transition, see Fig. \ref{scheme}a. The resulting emission on the $\ket{e}\leftrightarrow\ket{s}$ transition, which we denote anti-Stokes photons, is detected using the same set-up as for the Stokes photons, as shown in Fig. \ref{scheme}c. In the experimental histogram shown in Fig. \ref{scheme}b, the anti-Stokes detection probability is slightly larger than in the Stokes gate due to different sources of noise (see Supplemental Material).

By measuring the coincidences between the Stokes and anti-Stokes photons, we can calculate the second-order cross-correlation function $g_{S,aS}$ of the photon pair \cite{Chou2004}. If the average photon number is much less than 1 per mode, then $g_{S,aS}$ can be written as

\begin{equation}
g_{S,aS} \simeq \frac{p_{S,aS}}{p_{acc}},
\end{equation}

\noindent where $p_{S,aS}$ is the coincidence detection probability and $p_{acc}$ is the accidental coincidence probability due to uncorrelated photons. The AFC-based DLCZ scheme is producing specific time correlations due to the AFC rephasing mechanism. The read pulse will lead to a collective optical excitation that will continue to undergo inhomogeneous dephasing, due to the phase factor $\exp(-i 2\pi\delta_j (T_S+t'))$ of each atom $j$ where $t'$ is the time elapsed after the read pulse. The periodic AFC structure will ensure, however, that all atoms are in phase when $T_S+t' = 1/\Delta$, such that the emission probability of an anti-Stokes photon is strongly peaked for $t' = 1/\Delta-T_S$. 

In Fig. \ref{gsi}a we show the Stokes/anti-Stokes coincidence histogram as a function of the delay $\tau=T_{S}+T_{aS}$, where $T_{aS}=T_{\rm spin}+t'$. As expected there is a strong temporal correlation at $\tau = T_{\rm spin}+1/\Delta$ = 1020 $\mu$s. The triangular background of accidental coincidences stems from the convolution of the two square-shaped detection gates, see Fig. \ref{gsi}a. To calculate $p_{acc}$ for a given delay $\tau$ we need to consider all combinations of Stokes and anti-Stokes detections that could lead to an accidental coincidence at $\tau=T_{S}+T_{aS}$. By normalizing with this triangular-shaped $p_{acc}$ function, we obtain the cross-correlation function $g_{S,aS}$, shown in Fig. \ref{gsi}b. For the central bin this correlation function peaks at the value of 4.2 $\pm$ 0.5. The coherence time of the Stokes/anti-Stokes photon pair with respect to the detection gate shows that correlations in more than 10 modes are generated, see Fig. \ref{gsi}b.

In order to verify the non-classical nature of the Stokes/anti-Stokes correlations, one can use that for classical fields the cross-correlation function is bounded by the Cauchy-Schwarz inequality $R = g_{S,aS}^2 / (g_{S,S} \cdot g_{aS,aS}) \leq 1$ \cite{Chou2004}, where $g_{X,X}$ ($X=S$ or $aS$) are the auto-correlation functions for each field. 

We performed a simultaneous measurement of both the auto- and cross-correlation functions using $T_{spin}$ = 500 $\mu$s and $p_S = 0.002$ with $p_S$ the probability to emit at least one Stokes photon.
We obtained $g_{S,S} = 1.86 \pm 0.4$, $g_{aS,aS} = 1.96 \pm 0.36$ and $g_{S,aS} = 3.24 \pm 0.43$. The given results are calculated using a bin size of 100 ns. The violation of the Cauchy-Schwarz inequality $R=2.88^{+2.9}_{-1.4} > 1$ clearly demonstrates the non-classical nature of the Stokes/anti-Stokes correlations.

\begin{figure}
\begin{center}
\includegraphics[width=8cm]{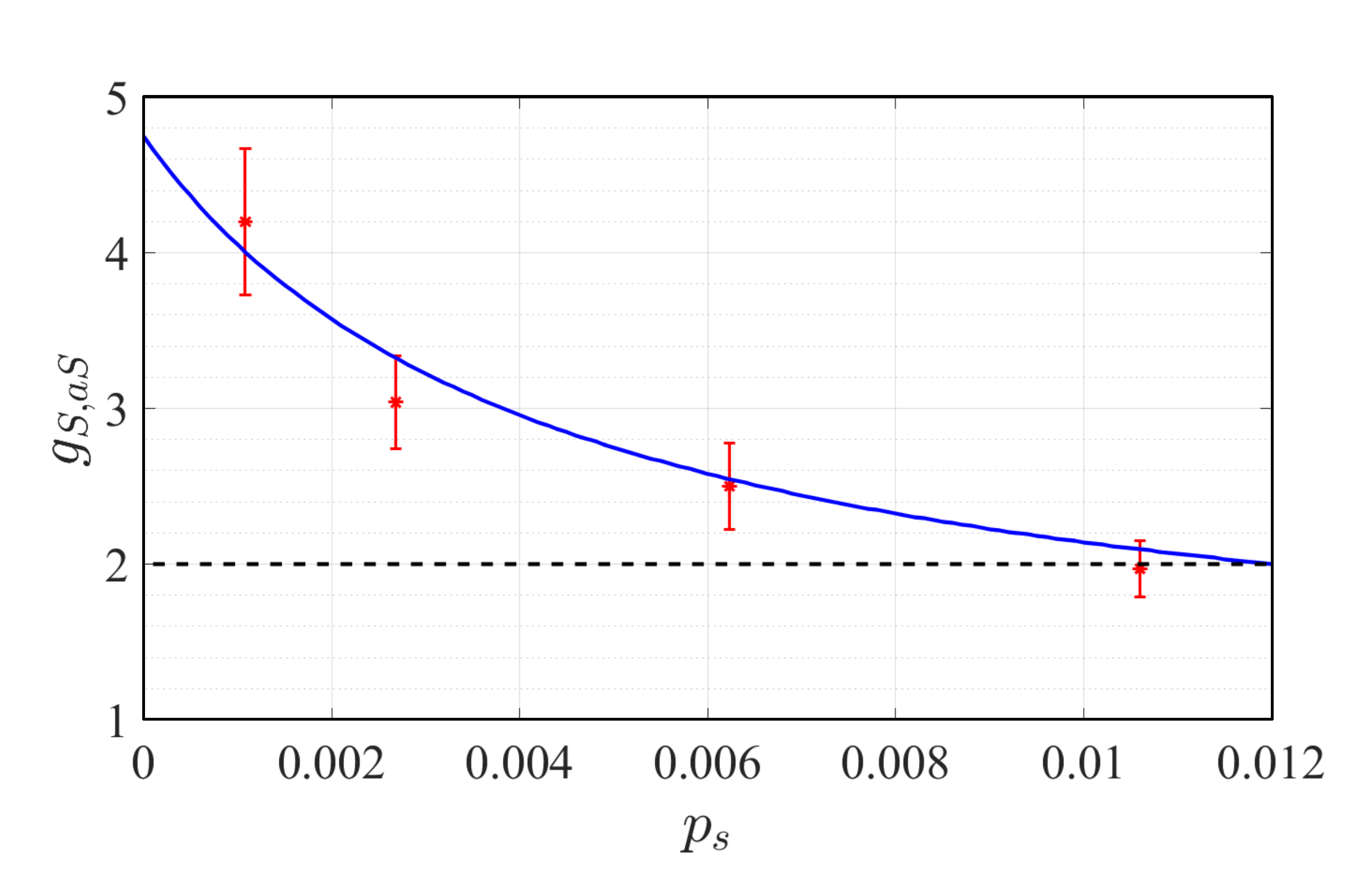}
\caption{Measured cross-correlation parameter $g_{S,aS}$ (red stars) as a function of the Stokes probability $p_S$, for the central bin of 100 ns at $\tau=T_{S}+T_{aS}$ shown in Fig. \ref{gsi}b. The theoretical model, see Eq. \ref{gsiformula}, is in close agreement with the experimental data (blue line). The model parameters were $p_n=0.12\%$, $\beta=0.27$ and $\eta_R=0.45\%$, all given for a bin size of 100 ns. Note that the parameters were measured independently and not fitted (see text for details). For an ideal two-mode squeezed state $g_{S,aS}$ is bounded by 2, however, the observed violation of the Cauchy-Schwarz inequality presents a more stringent test of the non-classical nature of the correlations.}
\label{gsi_ps}
\end{center}
\end{figure}

We can analyze the generation of quantum correlations in terms of a photon pair source. As a model we use the two-mode squeezed vacuum state \cite{Duan2001,Chou2004}, which is fully described by $p_S$. We also include a finite read-out efficiency of the anti-Stokes photon $\eta_R$, conditioned on the detection of a Stokes photon, and a source of uncorrelated noise in the anti-Stokes mode with probability $p_n$. Using this simple model the $g_{S,aS}$ function is given by (see Supplemental Material):

\begin{equation}
g_{S,aS}=1+\frac{\eta_R}{(\eta_R+\beta) p_S+p_n}.
\label{gsiformula}
\end{equation}

\noindent The term $\beta p_S$ accounts for spontaneous emission in the anti-Stokes gate that depends on the write pulse. Here $\beta$ is the fraction of population in the excited state at the time of the anti-Stokes emission, with respect to the initial population after the write pulse. We emphasize that $p_n$ and $\eta_R$ could be calculated independently from the experimental data and $\beta$ can be calculated using known experimental and atomic parameters of Eu$^{3+}$:Y$_2$SiO$_5$ (see Supplemental Material). As shown in Fig. \ref{gsi_ps}, the theoretical model predicts very well the cross correlation measurement as a function of the Stokes emission probability $p_S$, without any further adjustment of the parameters.

The measured anti-Stokes efficiency of $\eta_R=0.45\%$ used in the model shown in Fig. \ref{gsi_ps} refers to the efficiency in a temporal bin of 100 ns, which is much shorter than the coherence time of the Stokes/anti-Stokes photon pair $\tau_c$ = 0.41 $\mu$s (see Fig. \ref{gsi}). The temporal bin duration of 100 ns was found to represent the best compromise between the read-out noise $p_n$ and the efficiency $\eta_R$, leading to the highest cross-correlation $g_{S,aS}$. But if we consider a temporal bin of duration 2$\tau_c$, then the measured anti-Stokes read-out efficiency is 2.5 $\pm$ 0.3 $\%$. Our theoretical modeling based on the theory presented in \citep{Sekatski2011} gives an estimated efficiency between 5-10\%, depending on assumptions in the model parameters. The discrepancy with the experiment is currently not understood, but could be due to experimental issues, such as phase matching problems, or due to wrong assumptions in the theoretical modeling. Moreover, the measured cross-correlation was the same for the two storage times used in this work, $T_{spin}$ = 0.5 and 1 ms, while based on previous measurements \citep{Jobez2015} we would have expected an increased efficiency and cross-correlation function for the shorter spin-storage time. Future experiments should investigate the read-out efficiency in more detail, in particularly as a function of spin-storage time. An interesting approach in this context is to use a cavity-enhanced scheme, which resulted in an efficiency increase of a factor of 10 in a AFC spin-wave storage experiments \cite{Jobez2014}. Also, Zhong et al. \citep{ZhouHuaLiuEtAl2015} have shown that a coherence time of up to 6 hours can be achieved in this crystal, by applying a precisely aligned and strong magnetic field.

In conclusion, we have experimentally demonstrated a new multimode DLCZ scheme for generating non-classical correlations between a single photon and a collective spin excitation in a solid-state ensemble. The key feature of the scheme is the use of an atomic frequency comb to control the inhomogeneous dephasing, which allowed us to generate spin-photon correlations in at least 12 temporal modes. Even more modes could in principle be generated by increasing the delay $1/\Delta$, or by increasing the bandwidth of the AFC. Neither of these two parameters have reached their intrinsic limit in the specific case of $^{151}$Eu$^{3+}$:Y$_2$SiO$_5$ \cite{Jobez2016}. The read-out efficiency was low in the present demonstration, which in combination with the technical photon noise source limited the observed cross correlation function. This is not a fundamental issue, however, as noise can be reduced by further optimizing the spectral filtering elements and the efficiency can be boosted by a cavity approach, as we recently demonstrated \cite{Jobez2014}. The ability to generate multimode quantum correlations with long storage time is a key requirement for realizing quantum repeaters for long-distance quantum communication. Therefore we believe our results represent an important step for realizing a quantum repeater based on solid-state ensembles.\\

The authors thank Genko Genov, Nuala Timoney and Anthony Martin for useful discussions, as well as Claudio Barreiro for technical support. We acknowledge funding from the Swiss programme National Centres of Competence in Research (NCCR) project Quantum Science Technology (QSIT), EU’s H2020 programme under the Marie Skłodowska-Curie project QCALL (GA 675662) and EU's FP7 programme under the ERC AdG project MEC (GA 339198).

\nocite{Supp}
\nocite{Tian2011}
\bibliography{AFC-DLCZ.bib}
\bibliographystyle{apsrev4-1}

\newpage
\section{Supplemental Material}
\section*{The $^{151}$Eu$^{3+}$:Y$_2$SiO$_5$ crystal}
We use a Y$_2$SiO$_5$ crystal doped with $^{151}$Eu$^{3+}$ ions at a concentration of 0.1$\%$. We work on the ${}^7\text{F}_0\leftrightarrow{}^5\text{D}_0$ transition at 580.04 nm of site 1 \cite{Koenz2003}. The electronic ground and excited states each consist of three nuclear quadrupole states $\ket{\pm1/2}$, $\ket{\pm3/2}$ and $\ket{\pm5/2}$. The $\ket{g}$ and $\ket{s}$ states correspond to the $\ket{\pm1/2}$ and $\ket{\pm1/2}$ states respectively, in the electronic ground state. The $\ket{e}$ state corresponds to $\ket{\pm5/2}$ in the electronic excited state. The crystal is cool down to about 4 K in a cryogen-free low-vibration cooler with optical access.  At this temperature we measure an optical coherence time of 400 $\mu$s and spin coherence time of about 1 ms.

\section*{Atomic frequency comb}
 The AFC period is set to $1/\Delta$ = 20 $\mu$s in this experiment, which is a compromise in terms of number of temporal modes and the rephasing efficiency of the AFC comb. The comb is prepared on the $\ket{e}\leftrightarrow\ket{s}$ transition, which has the highest transition strength. Using the double-pass configuration described in Fig. 1c, we obtain an effective optical depth \cite{Afzelius2009a} of the comb of about $\tilde{d} \approx 1$. The AFC preparation method is described in detail in Ref. \cite{Jobez2016}.

\section*{Coherent spin control.}
 In solid-state ensembles, inhomogeneous broadening of the spin transition limits the storage time. The broadening of our particular $\ket{g}\leftrightarrow\ket{s}$ transition at 34.5 MHz is about 27 kHz. In previous work we have shown that we can extend the storage time from about 10 $\mu$s to around 1 ms using a spin echo sequence, while suppressing to an acceptable level the additional noise due to the spin echo sequence \cite{Jobez2015}. With respect to Ref. \cite{Jobez2015} we here employ a new sequence with an odd number of pulses. It consists of three adiabatic pulses frequency chirped over 100 kHz with full-width at half-maximum (FWHM) durations of 45, 90 and 45 $\mu$s and relative phases of 0, $\pi/2$ and 0 (XYX sequence). The RF field is created using a 6 turns coil wrapped around the crystal, coupled to a high-power RF amplifier ($>$50 W) using an impedance-matched circuit. The maximum Rabi frequency is 60 kHz. The spin storage time is set to $T_{\rm spin}=1$ ms, except for the Cauchy-Schwarz inequality measurement where it is set to $T_{\rm spin}=0.5$ ms. Note that the effect of the RF excitation on the spatial phase matching can be neglected, as the spin wavelength of 8 m is much longer than the crystal length (1 cm).

\section*{Write and read pulse parameters}
The read pulse should ideally be a population inversion pulse. To achieve a high degree of inversion over a large bandwidth, we employ a frequency chirped hyperbolic-square-hyperbolic pulse \cite{Tian2011} with a total duration of 8 $\mu$s and a bandwidth of 2 MHz. The read pulse transfer efficiency is about $\eta_T \approx 75\%$. The write pulse should have identical duration and chirp width as the read pulse, but with an opposite chirp direction. In this way one can ensure that all ions within the excitation bandwidth are rephased simultaneously thanks to the AFC structure. Note also that the write pulse is weaker, as it should excite a small fraction of the ions in $\ket{g}$. Furthermore, the bandwidth of the Stokes photons is given by the bandwidth of the write pulse (2 MHz), as it is narrower than the AFC bandwidth (5 MHz).

\section*{Experimental sequence}
The experimental sequence is divided into two parts. First we perform a 575 ms long preparation sequence, in which the ensemble is shaped into the atomic frequency comb structure over a 5 MHz bandwidth. Then we perform the experimental sequence composed by the write, RF and read pulses. After each sequence, a 10 ms repump pulse is applied on the $\ket{g}\leftrightarrow\ket{e}$ transition to re-polarize the spins to $\ket{s}$ and a RF inversion pulse is applied to re-initialize the ions into $\ket{g}$. This sequence is then repeated 14 times in order to increase the rate.

\section*{Noise filtering}
To detect the Stokes/anti-Stokes photons, filtering techniques have to be employed in order to remove as many noise photons as possible. These can be photons scattered from the write and read pulses, but also photons from fluorescence or free-induction decay emissions from the europium ions \cite{Jobez2015}. The angle between the write/read and Stokes/anti-Stokes modes provides a spatial filtering. The time-separation between the Stokes (resp. anti-Stokes) photons and the write (resp. read) pulses also results in a time filtering. Finally we apply frequency filtering using a Fabry-Perot cavity (finesse 200 and bandwidth of 2.5 MHz) that is frequency-locked to the Stokes/anti-Stokes frequency.

\section*{Theoretical model for cross-correlation}
A DLCZ-type source is equivalent to a two-mode squeezed vacuum state as generated, for instance, by a spontaneous down conversion process in non-linear optics \cite{Duan2001, Chou2004}. Such a state can be written as
\begin{equation}
\ket{\psi}=\sqrt{1-p_S}\sum_{n=0}^{\infty}p_S^{n/2}\ket{n}_S\ket{n}_{aS},
\end{equation}
where $p_S$ is the probability to emit at least one Stokes photon. We take this probability to be equal to the average photon number, due to the very low probability to detect more than one photon in each mode. To model our experiment we also include the read-out efficiency of the anti-Stokes photon $\eta_R$ and a source of uncorrelated noise in the anti-Stokes mode with probability $p_n$. This noise is produced by the read out pulse and the spin-echo sequence, as discussed in detail in Ref. \cite{Jobez2015}. Using this simple model we obtain the formula given in the main text: 

\begin{equation}
g_{S,aS}=1+\frac{\eta_R}{(\eta_R+\beta) p_S+p_n}.
\label{gsiformula}
\end{equation}

The term $\beta p_S$ accounts for spontaneous emission in the anti-Stokes gate due to population in the excited state induced by the write pulse. This population stems from two sources: 1) ions that remain in $\ket{e}$ after the read pulse, which is proportional to $1-\eta_T$ times the branching ratio to the $\ket{s}$ state. 2) ions that have decayed to the state holding the excitation (i.e. $\ket{s}$ during $\frac{T_{\rm spin}}{2}$ and $\ket{g}$ during $\frac{T_{\rm spin}}{2}$) that is subsequently excited by the read pulse. For the latter source we also need to consider the branching ratios $\gamma_{es}$ and $\gamma_{eg}$. $\beta$ is thus given by :

\begin{equation}
\beta=(1-e^{-\frac{T_{spin}}{2T_1}})(\gamma_{es}+e^{-\frac{T_{spin}}{2T_1}}\gamma_{eg})\eta_T+(1-\eta_T)\gamma_{es}e^{-\frac{T_{spin}}{T_1}}.
\label{betaformula}
\end{equation}

For the crystal used in the experiment we know $T_1$=1.9 ms, $\gamma_{es}$=0.75 and $\gamma_{eg}$=0.2 and given $T_{spin}$=1 ms, we estimate that $\beta = 0.27$. 

The noise probability $p_n$ can be measured independently by not applying the write pulse, while applying the RF and read pulses.

The readout efficiency is calculated from experimental results as

\begin{equation}
\eta_R=\frac{p_{coinc}-p_{acc}}{p_S}.
\label{etaformula}
\end{equation}

\end{document}